\documentclass[aps,prl,twocolumn,showpacs,preprintnumbers,amsmath,amssymb,10pt]{revtex4-1} 
\usepackage{amsmath,graphicx,subfig,xcolor,verbatim}
\usepackage{amsfonts}
\usepackage{diagbox}
\usepackage{tikz}
\usepackage{placeins} 

\newcommand{\be}{\begin{equation}}
\newcommand{\ee}{\end{equation}}
\newcommand{\bea}{\begin{eqnarray}}
\newcommand{\eea}{\end{eqnarray}}
\newcommand\restr[2]{{\left.\kern-\nulldelimiterspace#1\vphantom{\big|}\right|_{#2}}}

\newcommand{\beq}{\begin{equation}} 
\newcommand{\eeq}{\end{equation}}

\def\leq{\leqslant}

\begin{document}
\title{On the Ising/tricritical-Ising interface: \\ From $\mathcal{W}_3$ symmetry to Rydberg atoms}
\author{Ant\'onio Antunes$^Z$, Junchen Rong$^X$} 
\affiliation{
$^Z$Laboratoire de Physique de l'\'Ecole Normale Sup\'erieure, Universit\'e  PSL, CNRS, Sorbonne Universit\'e, Universit\'e  Paris Cit\'e, 24 rue Lhomond, F-75005 Paris, France \\
$^X$CPHT, CNRS, \'Ecole Polytechnique, Institut Polytechnique de Paris, Palaiseau, France}

\begin{abstract}
We consider interfaces between critical spin-chains in different universality classes, described in the continuum limit by defect/interface conformal field theory (DCFT/ICFT).
We focus on a conformal interface between the Tricritical Ising (TIM) and the Ising CFT exploring the possibility of its experimental realizations in the context of Rydberg atom arrays.  
Our analysis emphasizes non-invertible symmetries and consistency under modular transformations, and uses defect couplings and the defect spectrum -- including in the case of mixed boundary conditions -- to make sharp experimental predictions. 
The structure of the observables hinges on a newly discovered pattern of emergent $\mathcal{W}_3$ chiral symmetry for the Tricritical-Ising/Ising interface.
\end{abstract}
\maketitle
\nopagebreak

{\bf Introduction}
In 1989, Cardy solved the boundary conditions for many two-dimensional conformal field theories;
along the way, he invented boundary conformal field theory~\cite{CARDY1989581}.
Slightly later, the boundary conformal field theory approach was generalized to study defects of critical quantum systems, including the famous Kondo problem~\cite{Kondo,AFFLECK1990517,AFFLECK1991641,AFFLECK1991849} and the defects of the critical Ising model~\cite{Oshikawa:1996ww,OSHIKAWA1997533}, after applying the folding trick.

Critical quantum defects correspond to conformal interfaces~\cite{Bachas:2001vj,Quella:2002ct,Graham:2003nc,Frohlich:2006ch,Quella:2006de,Bachas:2007td,Kormos:2009sk,Bachas:2013nxa} in which the theories on the left and right are identical. 
Interfaces between different CFTs are theoretically more challenging.
Special attention has been paid to the so-called renormalization group (RG) interfaces, first considered in \cite{Brunner:2007ur,Douglas:2010ic} and systematically implemented in \cite{Gaiotto:2012np}, which realize the RG as a map between two halves of spacetime rather than a map between energy scales, see Fig.1-left. The RG flow induces a relation between UV and IR operators which is encoded in the data of the RG interface.
In particular, Gaiotto \cite{Gaiotto:2012np} considered flows between the series of 1+1d unitary minimal model CFTs $\mathcal{M}_{m,m+1}$ (first studied in \cite{Zamolodchikov:1987ti}) and showed that the coefficients of the interface state correctly capture the UV-IR map $\mathcal{M}_{m,m+1} \to \mathcal{M}_{m-1,m}$ to leading order in the large $m$ expansion \footnote{See \cite{Behan:2021tcn,Antunes:2022vtb,Antunes:2024mfb,Behan:2025ydd,Antunes:2025erb,Antunes:2026jjf} for recent applications of this idea.}. 
Using the state-operator correspondence in 2 dimensions makes it possible to study the same system on the cylinder, where the spatial manifold $S^1$ gets divided into two halves, with two interface insertions separating the UV and IR phases as depicted in Fig.1-right.
\begin{figure}[t]
\includegraphics[width=0.5\textwidth]{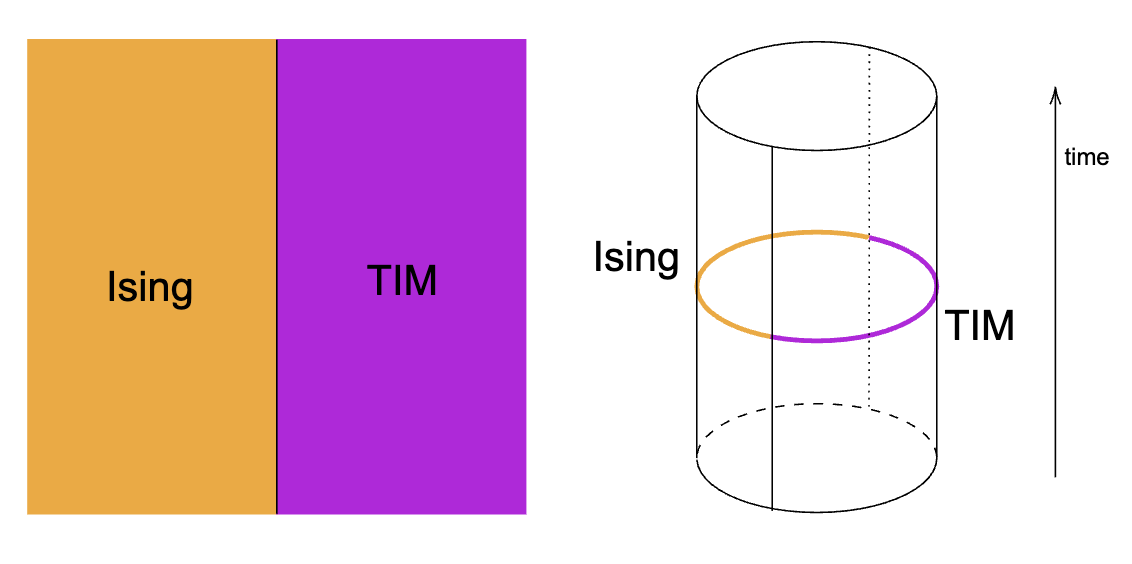}
\caption{A conformal interface on the plane becomes defects separating two different quantum critical phases on a circle, after a conformal transformation.}
\end{figure}

This framework is particularly appealing for non-perturbative studies at finite $m$, since $m=3$ corresponds to the critical Ising model, $m=4$  to the tricritical Ising model and $m=5$ to both the tetracritical Ising model and the critical 3-state Potts model, all of which can be easily realized with 1 dimensional quantum lattice models, especially when taking account their generalized symmetries \cite{Aasen:2016dop,Chang:2018iay,Aasen:2020jwb,Belletete:2020gst,Roy:2021jus}. 
In this work, we revisit the Ising-TIM interface and report on the discovery of new properties of this interface. 
We use the O'Brien-Fendley formulation \cite{OBrien:2017wmx}, which has manifest non-invertible Kramers-Wannier symmetry, to study the spectrum of a Hamiltonian in the presence of the interface~\footnote{ An alternative lattice model was implemented in~\cite{Cogburn:2023xzw} to test predictions of correlation functions by~\cite{Gaiotto:2012np}.}.
We will also discuss how to implement our interface using a simulator based on Rydberg-atom arrays.
In fact, the newly developed technology in~\cite{Sun:2026aqf} allows experimentalists to measure the energy spectrum of Rydberg atom chains. 
Studying quantum phase transitions and, moreover, measuring CFT properties using this platform is undergoing rapid development \cite{Keesling_2019,Scholl_2021,Ebadi_2021,Semeghini_2021,Fang:2024uyf}. 

Furthermore, we use the knowledge of the interface spectrum, along with modular invariance and modular bootstrap constraints, to reconstruct an exact interface state (up to a few unfixed signs). 
This state is constrained by a $\mathcal{W}_3$ chiral symmetry (generated by a spin-3 current), which is related to (but not the same as) the enhanced chiral symmetry in the tensor product of Ising and Tricritical Ising CFTs that Gaiotto used to guess his exact interface state \cite{Gaiotto:2012np}.
The precise relation to the construction of Gaiotto will be discussed in the upcoming work \cite{toappear}.

\noindent {\bf Erratum} 
In the previous version of this preprint, the interface constructed numerically  was misinterpreted as a new conformal interface between the Ising CFTs and tricritical Ising CFT. 
The correct statement is that the interface is precisely the RG domain wall constructed in~\cite{Gaiotto:2012np} in disguise. 
We will discuss this identification in detail in~\cite{toappear}.

{\bf Setup}
To study the interface, we use following Hamiltonian $H=H_L+H_R+H_b$, with a left ($H_L$) and right $(H_R)$ piece, supplemented by a localized interface interaction ($H_b$)
\begin{align}\label{hamiltonianInterface}
    H_L&= -J^*\left(\sum_{i=1}^{N/2-1}Z_i Z_{i+1} + \sum_{i=1}^{N/2}X_i\right) \\
    &+ K^*\sum_{i=1}^{N/2-2}\left( Z_i Z_{i+1} X_{i+2}+ X_i Z_{i+1} Z_{i+2}\right)\,, \nonumber\\
    H_R&= -\left(\sum_{i=N/2+1}^{N-1}Z_i Z_{i+1} + \sum_{i=N/2+1}^{N}X_i\right)\,,\\
    H_b&= h_b(Z_{N/2}Z_{N/2+1}+ Z_{N}Z_1)\,.
\end{align}
The left chain Hamiltonian $H_L$ was introduced in~\cite{OBrien:2017wmx}. 
If $H_L$ was defined on a chain with a periodic boundary condition, the system is invariant under the Kramers-Wannier (\textbf{KW}) duality, which acts as
\begin{equation}
    \textbf{KW}:\,X_i \to Z_i Z_{i+1}\,, \quad Z_i Z_{i+1}\to X_{i+1}\,,
\end{equation}
and in particular squares to a translation by a single lattice spacing.
When $K^*/J^*\approx0.428$, the large volume limit is described by the Tricritical Ising CFT~\cite{OBrien:2017wmx}.
Clearly, when $h_b=0$, the two chains are decoupled, i.e., the observables are determined by direct sums of Ising and tricritical Ising data in the presence of a conformal boundary condition.
We study the coupled chains with standard MPS/DMRG techniques, using the ITensor package~\cite{itensor}. 
We set the couplings $J^*=2/1.627$ and $K^*=(2/1.627)\times0.428$, which are chosen such that the speed of light, as measured from the finite volume scaling of the energy, matches on both sides.

\begin{figure}[ht]
\centering
\includegraphics[scale=0.35]{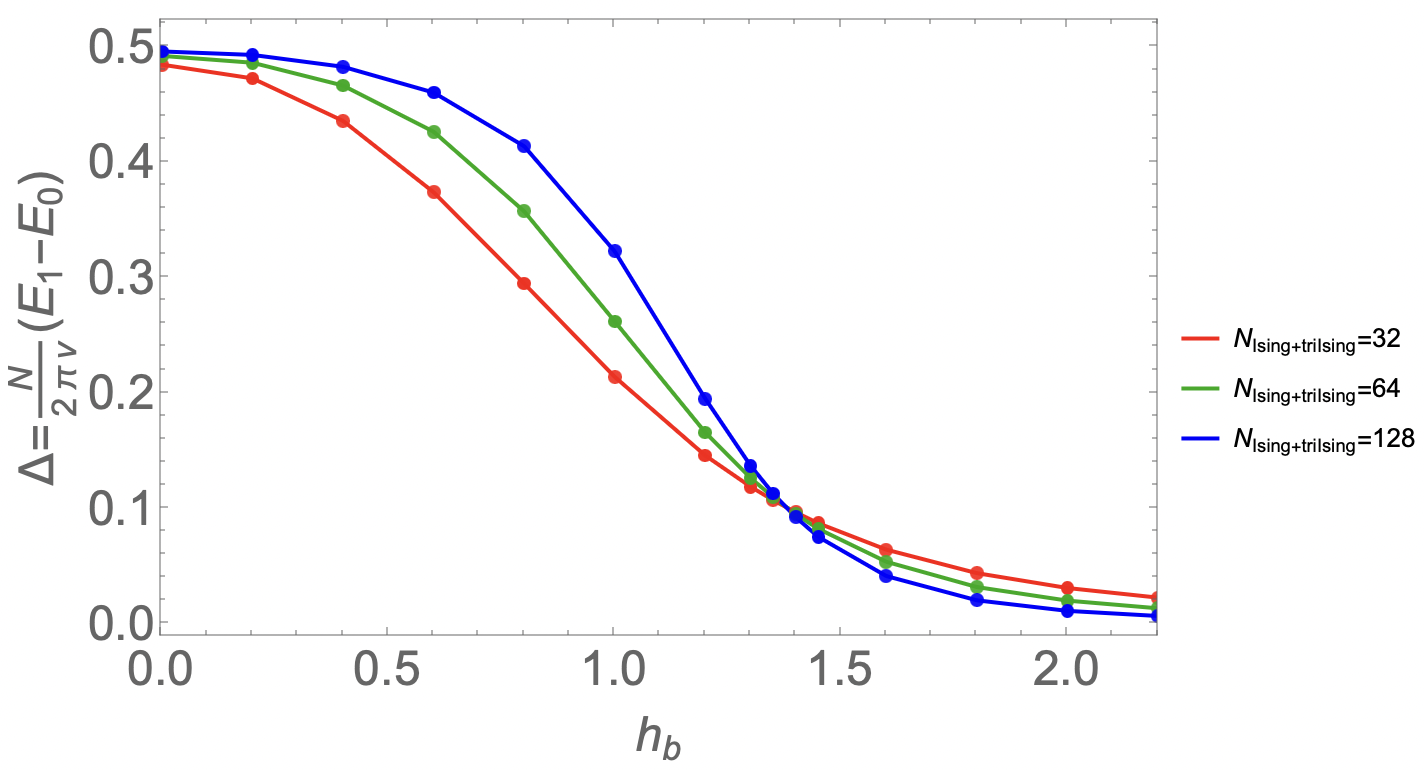}
\caption{The phase transition on the Ising/TIM interface. The plot is symmetric with respect to $h_b=0$.}\label{tri_Ising}
\end{figure}

{\bf Ising/Tricritical-Ising interface}
 To find non-trivial conformal interfaces, we scan over the interface coupling $h_b$ and plot the rescaled energy gap for different system sizes. We identify three points (up to the symmetry $h_b\leftrightarrow-h_b$) where crossing of different system sizes occurs, shown in Fig.~\ref{tri_Ising}. 
 
\begin{figure}[ht]
\centering
\includegraphics[scale=0.34]{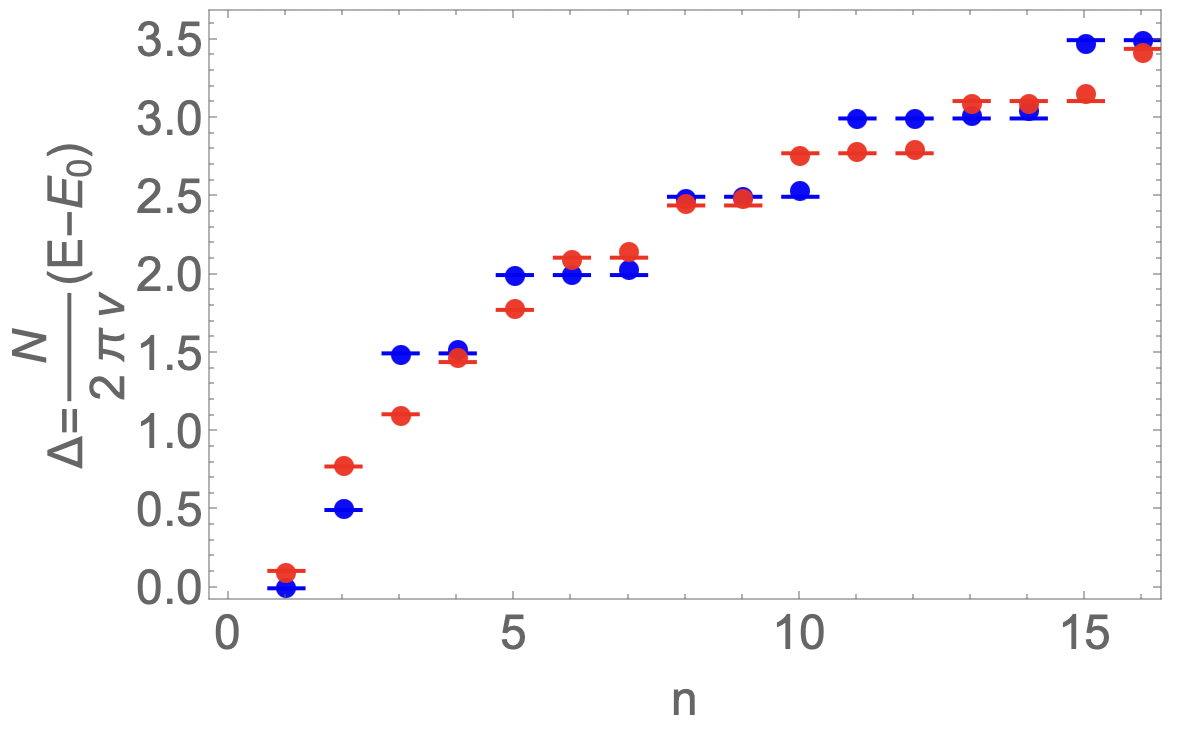}
\caption{The states of Ising/TIM interface at critical coupling, which corresponds to the open-string channel spectrum of the folded theory. 
The small horizontal lines are the theoretical prediction, and the dots are the numerical data.
The blue and red dots correspond to the $Z_2$ even and odd sectors, respectively.
The system size is $N_{\rm Ising}+N_{\rm TIM}=256$. 
}\label{tri_Ising_states}
\end{figure}

 Apart from the factorized free boundary conditions at $h_b=0$, we find a second factorized interface at $h_b\rightarrow+\infty$, corresponding to fixed spin (anti-ferromagnetic) boundary conditions, and a non-trivial interface at $h_b^*=1.379$. We will now focus our attention on this non-trivial interface which commutes both with the $\mathbb{Z}_2$ and $\textbf{KW}$ symmetries. We explicitly verify invariance under $\textbf{KW}$ transformations in Appendix D. To understand the $\mathbb{Z}_2$ structure, we compute the energy of several states and organize them into $\mathbb{Z}_2$ even and odd sectors, with the standard action of the $\mathbb{Z}_2$ symmetry in the TFIM. The results are presented in Fig.~\ref{tri_Ising_states}. 
For an open quantum spin chain at criticality, the energy is given by 
\begin{equation}\label{criticalspectrum}
    E_i=e_0N+\frac{\pi v}{N}(\Delta_i-\frac{c}{12})+\ldots,
\end{equation}
which relates the energy $E_i$ of the states to the scaling dimension $\Delta_i$ of CFT operators.
Here $e_0$ is the energy density, and $v$ is the so-called effective speed of light, which are non-universal constants. When we use the folding trick, the quantum spin chain~\eqref{hamiltonianInterface} with size $N$, can be viewed as an open chain with size $N/2$ and central charge $c=c_{\rm Ising}+c_{\rm TIM}=1.2$. From the spectrum, we can determine the scaling dimensions of operators in the open-string channel. Before studying the spectrum in more detail, let us note that the same interface can be obtained by deforming the tricritical Ising phase by an operator mapping to $\epsilon'$ in the continuum. We discuss the details of this alternative setup in Appendix C.

There are several interesting properties of the spectrum of this interface. First, we observe integer spacing consistent with conformal symmetry as well as a single relevant $\mathbb{Z}_2$ even operator $\Delta_+\approx0.5$ consistent with having to tune one coupling on the interface. There is also an approximate twofold degeneracy close to $\Delta\approx1.5$, as well as a subleading $\mathbb{Z}_2$-odd state around $\Delta_-'\approx0.77$. From the point of view of the $\rm{Vir}_{c=1/2}\times\rm{Vir}_{c=7/10}$ algebra, or of the enhanced symmetry described by Gaiotto and reviewed in Appendix A, there is no chiral weight ($h=h_{r,s}^{\rm{Ising}}+h_{r',s'}^{\rm{TIM}}$) in the correct ballpark to explain this state, which makes this boundary condition rather mysterious. We briefly sketch how such states could appear in the framework of \cite{Gaiotto:2012np} in Appendix B. To more clearly understand the origin of the puzzling state, one notes that there actually (at least) two rational theories with $c=1.2$. The tensor product of Ising and Tricritical Ising CFTs and the second $\mathcal{W}_3$ minimal model \cite{Zamolodchikov:1985wn,Fateev:1987vh} described by the coset
\begin{equation}
    (\mathcal{MW}_3)_{k=2}=\frac{su(3)_2\times su(3)_1}{su(3)_3}\,,
\end{equation}
which along with many weights which coincide with sums of Ising and Tricritical Ising values, further contains $h=\{1/9,7/9,13/9\}$ which can easily justify the numerically observed spectrum. We propose the following defect-channel/open-string partition function in terms of 
$\mathcal{W}_3$ characters,
\begin{align}
Z_{\rm{open}}(\tilde{q})&=\chi_{h=0}^{\mathcal{W}_3}(\tilde{q})+\chi_{h=1/2}^{\mathcal{W}_3}(\tilde{q})+2\chi_{h=2}^{\mathcal{W}_3}(\tilde{q}) \nonumber\\
&+\chi_{h=1/9}^{\mathcal{W}_3}(\tilde{q})+\chi_{h=7/9}^{\mathcal{W}_3}(\tilde{q})+\chi_{h=13/9}^{\mathcal{W}_3}(\tilde{q})\,.
\end{align}
Apart from matching the numerically observed spectrum, we will provide additional consistency checks below. First, we note that the $\mathcal{W}_3$ algebra must be generated by a spin-3 current that we identify with \cite{Crnkovic:1989ug}
\begin{equation}\label{spin3current}
    W(z)=\psi(z)\partial_z G(z)- 3\partial\psi(z) G(z).
\end{equation}
where the fields $\psi$ and $G$ both comes from $\mathbb{Z}_2$ twisted sectors: The $\mathbb{Z}_2$ twisted sector of the Ising model contains a free fermionic field $\psi$ and 
the $\mathbb{Z}_2$ twisted sector of the tricritical Ising model contains a fermionic field $G$ with $h=3/2$, which is usually called the supercurrent. These operators are attached to a topological line which corresponds to the $[\mathbb{Z}_2^{\rm Ising}\times \mathbb{Z}_2^{\rm TIM}]^{\rm diag}$, but since our interface commutes with the $\mathbb{Z}_2$ symmetry, this topological line can end on the interface.
Bringing $W(z)$ to the interface, it becomes the $h=3$ operator on the interface that appears in the open string sector spectrum \footnote{Holomorphic operators which appear after folding accross the interface have been refered to as `phantom currents' in the recent literature \cite{Antinucci:2025uvj,Furuta:2025ahl}. In this case, we have a spin-3 phantom current and the analysis of \cite{Gaiotto:2012np} predicts one phantom current of spin $(1-r)^2/2$ for each operator $\phi_{(1,r)}^{\rm{IR}}\phi_{(r,1)}^{\rm{UV}}$.}.

We can now try to understand the closed string channel and the associated boundary state. Since our open-string channel spectrum is written as $\mathcal{W}_3$ conformal characters, we know how they transform under the S-transformation.
The closed string channel partition function is then
\begin{align}\label{partitionfunctionclosed}
&Z_{\rm{closed}}(q)=c_1 \chi_{h=0}^{\mathcal{W}_3}(q)+\frac{c_1}{2}\chi_{h=2}^{\mathcal{W}_3}(q)+ c_2 \chi_{h=3/5}^{\mathcal{W}_3}(q)\nonumber\\
&+2 c_2 \chi_{h=8/5}^{\mathcal{W}_3}(q)+c_2 \chi_{h=1/10}^{\mathcal{W}_3}(q)+\frac{c_1}{2} \chi_{h=1/2}^{\mathcal{W}_3}(q)\,,
\end{align}
with
$c_1=\sqrt{2-\frac{2}{\sqrt{5}}}$, and $ c_2=\sqrt{\frac{1}{10} \left(\sqrt{5}+5\right)}$. 
However, reconstructing the associated boundary state from the partition function \eqref{partitionfunctionclosed} is rather delicate. It is straightforward to check that the full $\rm{Vir}_{c=1/2}\times\rm{Vir}_{c=7/10}$ symmetry is not preserved as it is impossible to expand $Z_{\rm{closed}}(q)$ in terms of a sum of products of Ising and Tricritical Ising CFT characters with positive coefficients. In fact, the tensor product theory can be obtained from the $\mathcal{W}_3$ minimal model by gauging an invertible symmetry \footnote{To obtain the $\mathcal{W}_3$ theory from the tensor product is a more subtle question (which involves gauging non-invertible symmetries) which we will address in upcoming work.}. The $\mathcal{W}_3$ primaries relevant to our discussion are summarized in Table~\ref{W3primaries11}.
\begin{table}[h]
\centering
\begin{tabular}{|c|c|c|c|c|c|c|c|c|}
\hline
  & $1$ & $D$ & $D^{\dagger}$ & $Z$ & $Z^{\dagger}$ & $X$ & $E$  & $F$ \\ \hline
$h$ & 0   & 2   & 2             & 3/5 & 3/5           & 8/5 & 1/10 & 1/2 \\ \hline
$w$ & 0   & $6q$   & $-6q$            & $q$ & $-q$           & 0 & 0 & 0\\\hline
\end{tabular}
\caption{The $\mathcal{W}_3$ primaries. Here $q=2 \sqrt{\frac{2}{35}}$.}\label{W3primaries11}
\end{table}
Here $h$ is, as usual, the eigenvalue of $L_0$, and $w$ is the eigenvalue of $W_0$, which flips sign under charge conjugation.
In Appendix E we discuss the effect of gauging in more detail, but as usual, it amounts to projecting out non-invariant states and adding invariant states from the twisted sector.
One can figure out what the $\mathcal{W}_3$ primaries correspond to in the Vir$\times$Vir theory.
In particular, states with an odd number of $W_n$'s will be projected out when gauging the complex conjugation symmetry. 
However, when we consider the $\mathcal{W}_3$ Ishibashi states \cite{Honecker:1992qf,Fuchs:1998xy,Fuchs:1999zi,Caldeira:2003yr}
\begin{equation}     |h;w\rangle\rangle_{\pm}=\sum_{N=0}^{\infty}\sum_{j=1}^{d(N)} |h;w;N,j\rangle_L \otimes Q{|h;\pm w;N,j\rangle}_R\,,
\end{equation}
the $W$ generators are always paired up, appearing an even number of times. 
Therefore, whether the states are gauged away or not depends fully on the primary. 
This explains the appearance of $\mathcal{W}_3$ characters in the closed-string channel \eqref{partitionfunctionclosed}. 
There are two types of $\mathcal{W}_3$ Ishibashi states. For the `untwisted' Ishibashi state $|h;w\rangle\rangle_+$, $Q=1$. 
For the `twisted' Ishibashi states $|h;w\rangle\rangle_-$, on the other hand, the operator $Q$ implements the outer automorphism $W_{-k}\rightarrow-W_{-k}$.
They satisfy twisted Cardy conditions described in Appendix E. 
All in all, this gives the following ansatz for the boundary state 
\begin{align}\label{bondaryansatz}
&|B\rangle_{\rm FM,\pm}=\sqrt{c_1}|1\rangle\rangle_{\mathcal{W}_3,\pm}+ s_1\sqrt{\frac{c_1}{2}}\frac{1}{\sqrt{2}} \big(|D\rangle\rangle_{\mathcal{W}_3,\pm}+c.c.\big)\nonumber\\&+s_2\sqrt{c_2} \frac{1}{\sqrt{2}}\big( |Z\rangle\rangle_{\mathcal{W}_3,\pm}+c.c.\big) +s_3\sqrt{2c_2}|X\rangle\rangle_{\mathcal{W}_3,\pm} \nonumber\\&+\sqrt{c_2}|\sigma,\sigma\rangle\rangle_{\mathcal{W}_3,\pm}+s_4\sqrt{\frac{c_2}{2}}|\sigma,\sigma'\rangle\rangle_{\mathcal{W}_3,\pm}\,, 
\end{align}
where the $s_i$ are unfixed signs at the moment. 
The definition of these Ishibashi states can be found in Appendix E.
All the states are invariant under complex conjugation gauging and under $\textbf{KW}$ duality (the Ising and Tricritical-Ising operators are in pairs that transform in the same way).
The sign of $|\sigma,\sigma\rangle\rangle_{\mathcal{W}_3}$ is a convention that depends on whether we introduce a ferromagnetic or antiferromagnetic interaction at the interface.
We will later show that one can fix $s_3=-1$, and more importantly, show that the $\mathcal{W}_3$ Ishibashi states have to be the twisted ones, by consistency under modular transformations for mixed boundary conditions: off-diagonal amplitudes in the closed-string channel must admit a positive-integer character decomposition in the open-string channel. 
The remaining signs $s_1$, $s_2$, and $s_4$, will be fixed in \cite{toappear} where it will also be shown that the state is actually the same as the one constructed in \cite{Gaiotto:2012np}.
 
\begin{figure}[ht]
\centering
\includegraphics[scale=0.35]{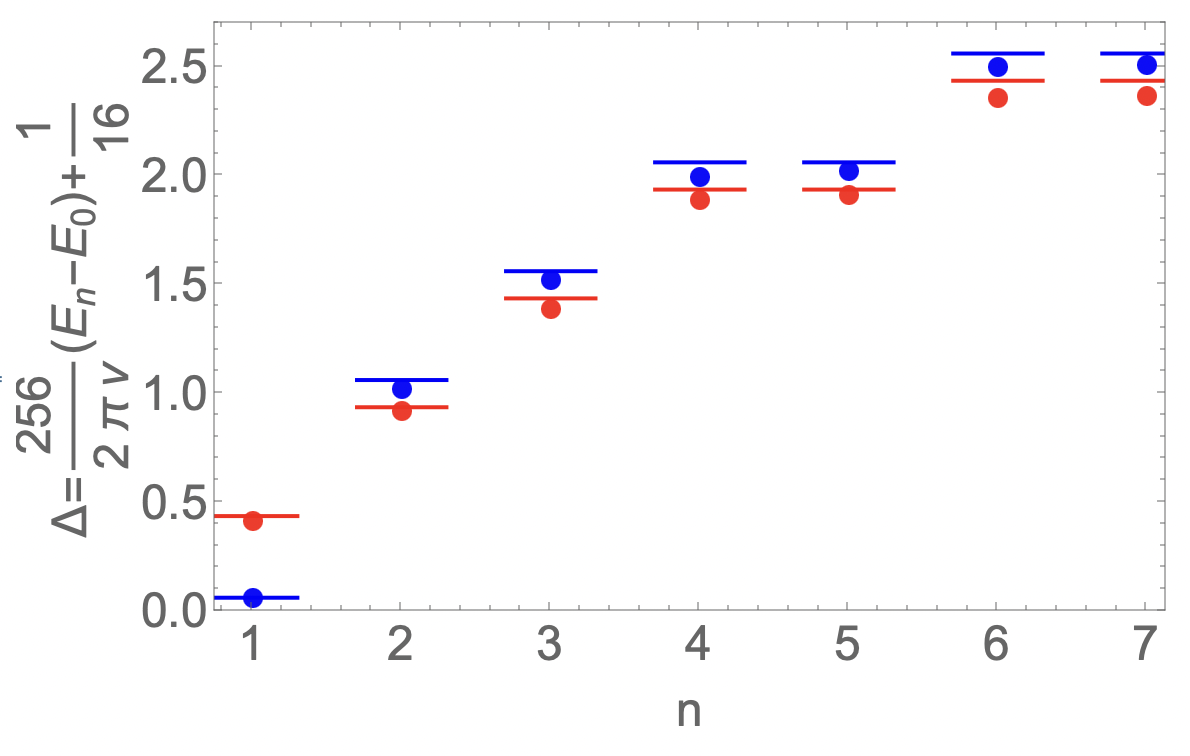}
\caption{The open-string sector spectrum of mixed boundary conditions. We apply the FM boundary condition on one side and the factorized boundary condition $C_1$ on the other side. The system size is $N_{\rm Ising}+N_{\rm TIM}=256$. 
The blue and red dots correspond to the $Z_2$ even and odd sectors, respectively. }\label{mixed_spectrum_FM_C1}
\end{figure}

\begin{figure}[ht]
\centering
\includegraphics[scale=0.35]{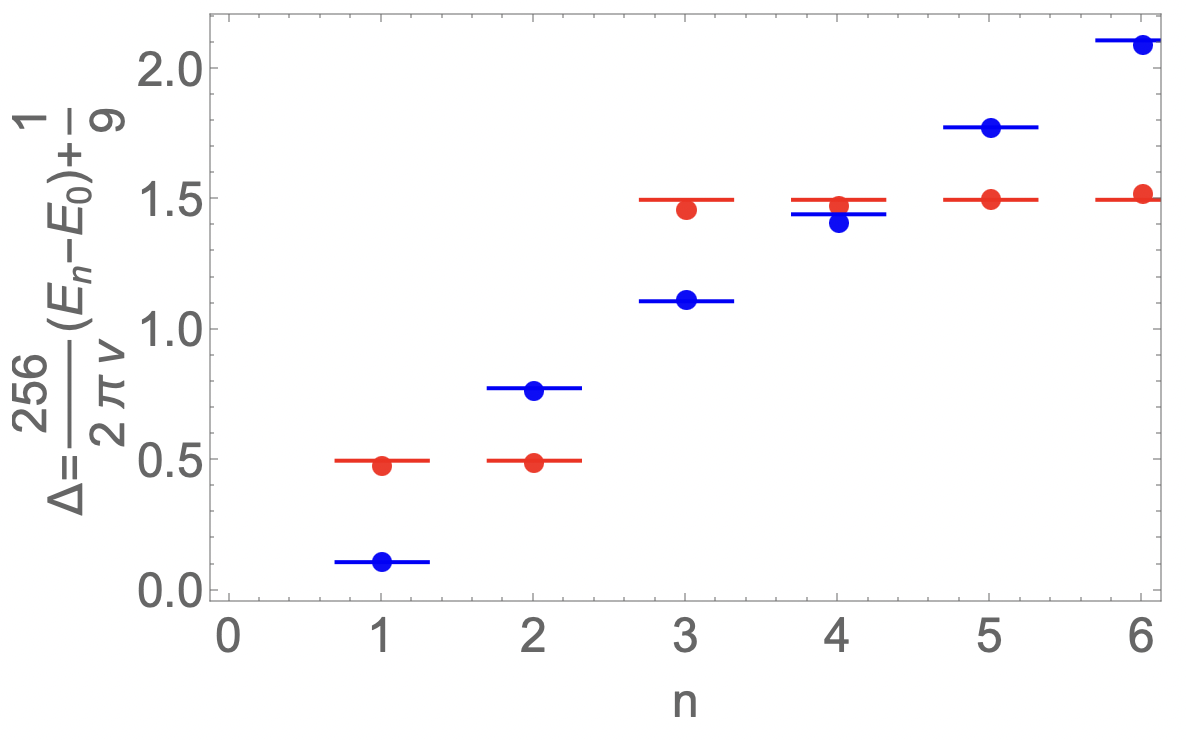}
\caption{The open-string sector spectrum of mixed boundary conditions. We take the FM boundary condition on one side and the AFM boundary on the other side. The system size is $N_{\rm Ising}+N_{\rm TIM}=256$. The blue and red dots correspond to the $Z_2$ even and odd sectors, respectively. }\label{mixed_spectrum}
\end{figure}

{\bf Mixed interfaces}
The simplest boundary states we can consider are tensor products of Cardy states of the Ising and Tricritical-Ising CFTs \cite{Cardy:1986gw,Cardy:1989irb,Lewellen:1991tb,Behrend:1999bn}. 
It turns out that out of the $3\times6=18$ Cardy states, two of them preserve the $\mathcal{W}_3$ symmetry 
 \begin{align}
 \label{mixedampli}
|C_1\rangle\equiv&|h=1/16\rangle_{\rm Cardy}^{\rm Ising} \otimes |h=7/16\rangle_{\rm Cardy}^{\rm TIM}\,,\nonumber\\
|C_2\rangle\equiv&|h=1/16\rangle_{\rm Cardy}^{\rm Ising} \otimes |h=3/80\rangle_{\rm Cardy}^{\rm TIM}\,,
\end{align}
with the decompositions in terms of $\mathcal{W}_3$ Ishibashi states given in Appendix D. 
Physically, the Cardy state $|h=1/16\rangle_{\rm Cardy}^{\rm Ising}$ is the free boundary condition of the Ising CFT, while the Cardy states $|h=7/16\rangle_{\rm Cardy}^{\rm TIM}$ and $|h=3/80\rangle_{\rm Cardy}^{\rm TIM}$ are the fixed vacancy and free boundary condition of the tricritical Ising CFT. 
Imposing integer positivity of the open string channel for the amplitudes
\begin{equation}
    Z_{B,C_i}(q)= {}_{\rm FM}\langle B| e^{-\beta H} |C_i\rangle_{}\,,\quad i=1,2\,,
\end{equation}
fixes $s_3=-1$, and selects the twisted version of the $\mathcal{W}_3$ states $|h,w\rangle\rangle_-$ in the ansatz \eqref{bondaryansatz} for the boundary state.
While the minus signs cancel in amplitudes between two twisted states, leading to ordinary $\mathcal{W}_3$ characters, the mixed amplitudes \eqref{mixedampli} contain overlaps between twisted and untwisted states, leading to the so-called twisted characters $\tilde{\chi}$, for example 
\begin{align}
        Z_{B,C_1}(q) &=\sqrt{c_1c_3}\tilde{\chi}^{\mathcal{W}_3}_{h=0}(q)-\sqrt{2}c_2\tilde{\chi}^{\mathcal{W}_3}_{h=8/5}(q)\,\nonumber\\
        &=\tilde{\chi}^{A_2^{(2)}}_{h=1/16}(\tilde{q})+\tilde{\chi}^{A_2^{(2)}}_{h=7/16}(\tilde{q}).
\end{align}
where $c_3=\sqrt{\frac{1}{10} \left(5-\sqrt{5}\right)}$. The modular transformation of twisted $\mathcal{W}_3$ characters is the sum of ordinary characters of the so-called affine twisted Lie algebra $A_2^{(2)}$, which we discuss further in Appendix E.
Since we impose a periodic boundary (untwisted) condition at one interface, and an anti-periodic (untwisted) boundary condition at the other interface, only half-integer modes $W_{k/2}$ of the spin-3 current are allowed. 
The boundary condition $|C_1\rangle$ corresponds to setting $h_b=0$ on one interface. We fixed the respective boundary conditions at each end of the chain and obtained the spectrum of Figure \ref{mixed_spectrum_FM_C1}, which is perfectly compatible with the half-spaced structure.
We can also consider the mixed Ferromagnetic/Anti-Ferromagnetic amplitude, where the AFM state is obtained by flipping the sign of the last two Ishibashi states in \eqref{bondaryansatz}, and physically by flipping the sign of the defect coupling. Analytically we obtain
\begin{align}
&Z_{\rm FM-AFM}(q)={}_{\rm FM}\langle B| e^{-\beta H} |B\rangle_{\rm AFM}\nonumber\\
&=\chi^{\mathcal{W}_3}_{1/9}(\tilde{q})+\chi^{\mathcal{W}_3}_{7/9}(\tilde{q})+\chi^{\mathcal{W}_3}_{13/9}(\tilde{q})+2\chi^{\mathcal{W}_3}_{1/2}(\tilde{q})\nonumber\,,
\end{align}
where $\chi(\tilde{q})$ are the characters in the open-string channel. We find a great match with the numerical spectrum presented in Figure \ref{mixed_spectrum}.

{\bf Experimental realization via Rydberg chains} We now discuss how to realize the conformal interface using a Rydberg atom array. 
The critical Ising CFT was realized in~\cite{Fang:2024uyf}, generalizing the 
earlier works on the non-critical Ising model~\cite{Labuhn:2016xba}. In~\cite{Vovrosh:2025mys}, the $E_8$ structure of the massive particles of the Ising model deformed by a magnetic field was experimentally observed.
Recently, the tricritical Ising CFT was also realized in~\cite{Sun:2026aqf}. 
In addition to that, there was a theoretical proposal in~\cite{Wang:2025nrd}, which realizes the tricritical Ising CFT using Rydberg atoms arranged on a two-ladder geometry. 
These pioneering works open the door for the experimental realization of the interface between the Ising CFT and the tricritical Ising CFT.
In a Rydberg atom array, the quantum Hamiltonian can be approximated by
\begin{equation}
H=\sum_{i,j} \frac{C_6}{r_{ij}^6} n_i n_j+\sum_{i} \left(\frac{\Omega}{2} X_i -\delta_i n_i\right)\,.
\nonumber
\end{equation}
Here, $n_i=\frac{1}{2}(1-Z_i)$ measures whether the atom is in a Rydberg state. $\Omega$ is the effective Rabi frequency and $\Delta$ is the detuning of the laser. 
In~\cite{Wang:2025nrd}, $C_6<0$ was achieved by using different atoms for neighboring rungs. 
We assume that the distance between atoms within the same rung is fixed. 
By adjusting the distance between neighboring rungs and the detuning $\delta$, we can place half of the chain at the Ising critical point and the other half at the tricritical Ising critical point~\cite{Wang:2025nrd}.
\begin{figure}[ht]
\centering
\includegraphics[width=0.5\textwidth]{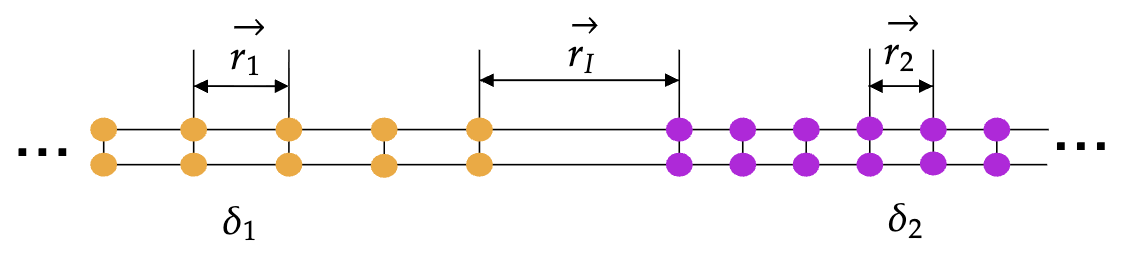}
\caption{Using Rydberg atoms to realize the conformal interface.}\label{interfaceRydberg}
\end{figure}
The interface coupling $h_b$ introduced in \eqref{hamiltonianInterface} can be tuned by changing the distance between the left and right chains.
In this setup, it is crucial to introduce spatially independent $\delta_i$. 
This can be achieved in experiments, as recently demonstrated in \cite{Manovitz_2025}, by adding an extra Spatial Light Modulator (SLM) to the experimental platform. 
The extra laser induces an AC Stark shift, which allows the experimentalists to achieve arbitrary detuning profiles for the atoms. 
Notice that the $\mathbb{Z}_2$ global symmetry that swaps the two chains is always preserved. 
This corresponds to the $\mathbb{Z}_2$ symmetry of the Ising model. 
Given the universality of critical systems, we expect the conformal interface in the setup above to be the same as that studied earlier.

If we imposed the free boundary conditions on the far left end of the Ising chain and on the far right end of the tricritical Ising chain, this corresponds to the $|C_1\rangle$ state in \eqref{mixedampli}. 
This is because both the boundary state of Ising $|h=1/16\rangle_{\rm Cardy}^{\rm Ising}$ and the boundary state of the tricritical Ising model $|h=7/16\rangle_{\rm Cardy}^{\rm TIM}$ are the most stable boundary conditions under the RG flow when the symmetry of $\mathbb{Z}_2$ is preserved.

To match the experimental spectrum with our CFT result, it is crucial to set 
\begin{equation}\label{speedmatch}
\frac{v_{\rm Ising}}{N_{\rm Ising}}=\frac{v_{\rm TIM}}{N_{\rm TIM}}
\end{equation}
As $v/N$ serves as the unit of energy for the critical system, see~\eqref{criticalspectrum}. 
Since the effective speed of light of the Ising phase and the TIM phase do not necessarily match, one needs to adjust $N_{\rm Ising}/N_{\rm TIM}$ to compensate this mismatch.

The possibility of simulating bilayered or even three-dimensional quantum systems using a Rydberg atom array has also been explored~\cite{Barredo_2018}. 
With these more advanced experimental platforms, one might be able to bend the ladderinto a ladder-ring geometry. 
This will allow us to introduce two non-trivial interfaces, therefore realizing a quantum system whose energy spectrum is given by our FM-FM and FM-AFM partition functions.

{\bf Discussion}
In this work, we considered a conformal interface between the Ising and Tricritical Ising CFTs, which we studied with a quantum lattice model. 
This allowed us to predict spectra for several configurations of boundary conditions which nicely match the numerical results.
We further suggested an experimental realization using arrays of Rydberg atoms in a ladder geometry. 

There are also many interesting generalizations waiting to be uncovered: Interfaces involving tetracritical Ising and the 3-state Potts CFT  which can be studied with the anyon chain formulation \cite{Feiguin_2007,PhysRevLett.101.050401,Antunes:2025huk} - lead to folded theories with higher central charge and could exhibit interesting symmetry enhancement patterns, or even so-called irrational boundary conditions \cite{Meineri:2025qwz}, with some of these also being realistically accessible experimentally. An orthogonal direction is to consider scenarios where the interfaces are not diametrically opposed on the circle, but form instead an angle $\theta$. The state-operator correspondence relates the energy of such Hamiltonians to scaling dimensions of edges/cusps \cite{Antunes:2021qpy,Diatlyk:2024zkk,Shimamori:2024yms,Shachar:2024cwk,Cuomo:2024psk} which are rich observables of great physical significance.

In addition to the Rydberg atom array, another notable experimental realization of the 2D Ising CFT and tricritical Ising CFT involved atoms adsorbed on 2D surfaces~\cite{PhysRevLett.44.152,WIECHERT1997164,PhysRevB.70.125410}.
It was also proposed theoretically that the tricritical Ising CFT can be realized using tunable Josephson junction ladders
~\cite{PhysRevLett.132.226502}, and the boundary mode of the Type IIID topological superconductor~\cite{Grover:2013rc}.
It will be interesting to explore the possibility of studying our conformal interface in these systems as well.

\vspace{10pt}

{\bf Acknowledgements}
We thank Jin Chen and Jiaxin Qiao for collaboration on related work and Costas Bachas, Liam Fitzpatrick, Davide Gaiotto and Giovanni Rizi for useful discussions. 
AA further thanks Andreia Gon\c{c}alves for continued inspiration. AA is funded by the European Union (ERC, FUNBOOTS, project number 101043588, PI Miguel Paulos). JR is funded by the European Union (ERC, QFTinAdS, project number 101087025, PI Balt van Rees).
Views and opinions expressed are however those of the authors only and do not necessarily reflect those of the European Union or the European Research Council Executive Agency. Neither the European Union nor the granting authority can be held responsible for them.
\bibliographystyle{apsrev4-1}
\bibliography{main}

\newpage
\onecolumngrid
\appendix

\begin{center}
\textbf{Appendix A: Brief review of Gaiotto's RG interface}    
\end{center}
In \cite{Gaiotto:2012np}, Gaiotto proposed an exact RG interface for consecutive unitary diagonal minimal models $\mathcal{M}_{m,m+1}^{\rm{UV}}\to\mathcal{M}_{m-1,m}^{\rm{IR}}$, which corresponds to a boundary condition of the folded theory $\mathcal{M}_{m,m+1}^{\rm{UV}}\otimes\mathcal{M}_{m-1,m}^{\rm{IR}}$. To do this he noted the following relation between cosets \cite{Crnkovic:1989ug}
\begin{equation}
    \frac{su(2)_k\times su(2)_1}{su(2)_{k+1}}\times\frac{su(2)_{k+1}\times su(2)_1}{su(2)_{k+2}}\simeq  \frac{su(2)_{k}\times su(2)_1\times su(2)_1}{su(2)_{k+2}}\,,
\end{equation}
where $k=m-2$ and the right-hand side makes it clear that an enhanced symmetry is present. Not only are there two stress-tensors, but there is an additional set of currents in different twisted sectors, i.e. attached to topological lines. In particular, the chiral operators
\begin{equation}
    \mathcal{J}_r=\phi^{\rm{IR}}_{(1,r)}\phi^{\rm{UV}}_{(r,1)}\,, \quad h_r= \frac{(r-1)^2}{2}\,,
\end{equation}
attached to the topolgical lines $\mathcal{L}^{\rm{IR}}_{(1,r)}\mathcal{L}^{\rm{UV}}_{(r,1)}$ are integer spin currents for odd $r$ and half-integer spin currents for even $r$. The bosonic sub-algebra generated by the integer spin currents is denoted by $\mathcal{B}$ and the full algebra including fermions is denoted by $\tilde{\mathcal{B}}$. In terms of the tensor-product Virasoro algebra $\mathcal{A}$, representations obey the following branching rules:
\begin{equation}
\label{branch}
    [\phi_{[t,d,\tilde{d},s]}^{\mathcal{B}}]= \sum_{r} [\phi_{t,r}^{\rm{IR}}]\times [\phi_{r,s}^{\rm{UV}}]\,,
\end{equation}
where $d,\tilde{d}$ label the two intermediate $su(2)_1$ representations. For each pair $t,s$ one has two possible choices for $d,\tilde{d}$, which can be further combined to form a representation of the fermionic algebra $\tilde{\mathcal{B}}$. For practical calculations it is useful to note the further identification with the tensor product of a super-symmetric minimal model and an Ising model
\begin{equation}
    \tilde{\mathcal{B}}\simeq \mathcal{S}\mathcal{M}_{k+1,k+3}\times\mathcal{M}_{3,4}\,,
\end{equation}
where the currents can be built out of the two stress-tensors, the supercurrent $G$ of the susy minimal model and the twisted sector fermion $\psi$ attached to the $\mathbb{Z}_2$ line in the Ising model. With all the notation in place we can finally write down the proposed boundary state
\begin{equation}
\label{RGState}
   |RG\rangle= \mathcal{I}_{\mathcal{B}\to \mathcal{A}} |\tilde{\mathcal{B}}\rangle_{\mathbb{Z}_2}=\sum_{t,s,r}^{s-t \in 2 \mathbb{Z}}\frac{\sqrt{S_{1,t}^{(k-1)} S_{1,s}^{(k+1)}}}{S_{1,r}^{(k)}} \,||\mathcal{A};t,r;r,s||t,s,\tilde{\mathcal{B}}\rangle\rangle_{\mathbb{Z}_2}\,,
\end{equation}
where the $\mathbb{Z}_2$ subscript denotes that the $\tilde{\mathcal{B}}$ Ishibashi/Cardy states are twisted under the $\mathbb{Z}_2$ symmetry acting on the Ising fermion $\psi$, $S_{r,s}^{(k)}$ is the modular S-matrix of the $su(2)_k$ WZW model and there is a projection via the topological interface $\mathcal{I}_{\mathcal{B}\to \mathcal{A}}$ from each $\tilde{\mathcal{B}}$ Ishibashi state to the corresponding states of the $\mathcal{A}$ theory, satisfying the branching rules \eqref{branch}. Crucially, these are not simply Ishibashi states of the $\mathcal{A}$ theory, because of the non-trivial $\mathbb{Z}_2$ action.

\begin{center}
\textbf{Appendix B: Defect channel of Gaiotto's RG interface}    
\end{center}
Given Gaiotto's RG state, we can now consider the closed string channel amplitude
\begin{equation}
  Z_{RG}(q)= \langle RG| q^{L_0 + \bar{L}_0+\frac{c}{12}}|RG\rangle\,,
\end{equation}
which can be evaluated using the construction
of the RG boundary state in terms of the $\mathbb{Z}_2$-twisted $\tilde{\mathcal{B}}$ states. Indeed, we have
\begin{equation}
      Z_{RG}(q)=\, _{\mathbb{Z}_2}\!\langle \tilde{\mathcal{B}}|(\mathcal{I}_{\mathcal{B}\to \mathcal{A}})^\dagger q^{L_0 + \bar{L}_0+\frac{c}{12}} \mathcal{I}_{\mathcal{B}\to \mathcal{A}}|\tilde{\mathcal{B}}\rangle_{\mathbb{Z}_2} =\,_{\mathbb{Z}_2}\!\langle \tilde{\mathcal{B}}| q^{L_0 + \bar{L}_0+\frac{c}{12}} (\mathcal{I}_{\mathcal{B}\to \mathcal{A}})^\dagger\mathcal{I}_{\mathcal{B}\to \mathcal{A}}|\tilde{\mathcal{B}}\rangle_{\mathbb{Z}_2}\,,
\end{equation}
where in the second equality we used that the interface is topological. We have therefore reduced the computation of the amplitude to determining the action of $(\mathcal{I}_{\mathcal{B}\to \mathcal{A}})^\dagger\mathcal{I}_{\mathcal{B}\to \mathcal{A}}$ on $|\tilde{\mathcal{B}}\rangle_{\mathbb{Z}_2}$. This turns out to be a rather delicate calculation. The topological interface is defined via \cite{Gaiotto:2012np}
\begin{equation}
  \mathcal{I}_{\mathcal{B}\to \mathcal{A}}   (\mathcal{I}_{\mathcal{B}\to \mathcal{A}})^\dagger= 1_{\mathcal{A}} +\eta_{\mathcal{A}}\,,
\end{equation}
which is essentially a statement that $\mathcal{B}$ is obtained by a $\mathbb{Z}_2$ gauging of $\mathcal{A}$, where the symmetry is realized as an ordinary Verlinde line. However, to calculate the partition function, we need the opposite operation, $(\mathcal{I}_{\mathcal{B}\to \mathcal{A}})^\dagger\mathcal{I}_{\mathcal{B}}$, which presumably gauges back a `quantum' $\tilde{\mathbb{Z}}_2$ symmetry to obtain the $\mathcal{A}$ theory from the $\mathcal{B}$ theory.
Therefore, we have schematically
\begin{equation}
      Z_{RG}(q)=\, _{\mathbb{Z}_2}\!\langle \tilde{\mathcal{B}}|q^{L_0 + \bar{L}_0+\frac{c}{12}} (1_{\mathcal{B}}+ \tilde{\eta}_{\mathcal{B}})|\tilde{\mathcal{B}}\rangle_{\mathbb{Z}_2} = \sum_i  \chi^{\mathcal{B}}_i(q) + \sum_j  \tilde{\chi}^{\mathcal{B}}_j(q)\,,
\end{equation}
which involves both ordinary character $\chi$ and characters twisted by the `quantum' $\tilde{\mathbb{Z}}_2$ symmetry $\tilde{\chi}$. After a modular transformation onto the open-string channel, the first set of characters will need to ordinary weights of the untwisted sector, while the second set can lead to the non-trivial $\mathcal{W}_3$ weights discussed in the main text.
The subtlety of the computation lies in the identification of the correct action of $\tilde{\mathbb{Z}}_2$ and the determination of the twisted characters $\tilde{\chi}$ which will be discussed in detail in the upcoming work \cite{toappear}.

\begin{figure}[ht]
\centering
\includegraphics[scale=0.5]{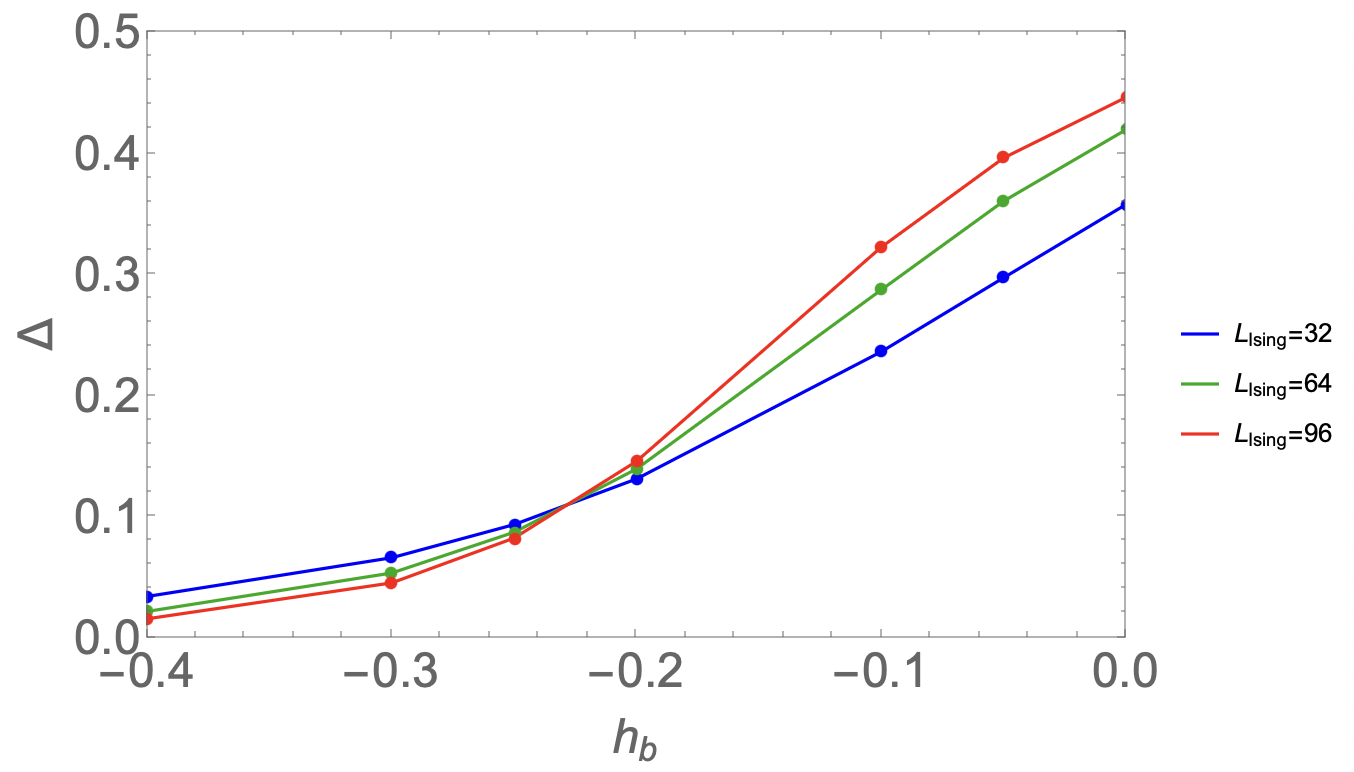}
\caption{The spectrum of the lattice RG interface.}\label{latticeRG}
\end{figure}

\begin{center}
\textbf{Appendix C: A lattice RG interface}    
\end{center}
 In this section, we argue that our lattice interface indeed corresponds to the RG domain wall in the continuum field theory.
In the continuum, the RG domain wall can be understood as 
\begin{equation}
S_{\rm TIM} +\int_{-\infty}^{\infty} dt~\int_0^{\infty} dx \, \epsilon'(x,t)\,. \nonumber
\end{equation}
Here $S_{\rm TIM}$ denotes the formal action of the tricritical Ising model. The $\epsilon'$ operator triggers an RG flow towards the Ising model; since it is KW invariant, the IR is conformal. 
On the lattice, we build the same RG interface.  The tricritical Ising spin chain $H_{\rm L}$, as defined in \eqref{hamiltonianInterface}, is deformed by the term 
\begin{equation}
H_{\rm deformation} = J \left( \sum_{i=1}^{N_I} Z_i Z_{i+1} +X_i\right) + J X_{N_I}  \nonumber
\end{equation}
on part of the chain. 
This triggers a flow to the Ising model on part of the chain. 
The interface is located at $i=1$ and $i=N_i$.
On the lattice, however, this deformation does not preserve the KW symmetry, because KW involves lattice translation. 
This can be seen more explicitly on the lattice operator level.
The operator 
$$
O_i = Z_iZ_{i+1} +X_i, 
$$
after KW transformation becomes 
$$
KW[O_i]=  X_{i+1}+Z_iZ_{i+1}.
$$
Therefore 
$$
KW[O_i]- O_i = X_{i+1}-X_i \sim B \partial_x \epsilon(x).
$$
Here $\sim$ means the identification of the CFT operator in the continuum limit. We can also make the following identification,
$$
O_x \sim A - \frac{B}{2} \partial_x \epsilon(x) +C \epsilon'(x)\,,
$$
where $A$ and $B$ are non-universal constants. 
The deformation is then,
\begin{align}
H_{\rm deformation}\sim&  \int_{1}^{N_I} dx \epsilon'(x) - \frac{B}{2} \int_{1}^{N_I} dx \partial_x \epsilon(x)+  B  \epsilon(x=N_I)\nonumber\\
\sim & \int_{1}^{N_I} dx \epsilon'(x) + \frac{B}{2}\epsilon(x=1)+ \frac{B}{2}\epsilon(x=N_I). \nonumber
\end{align}
We immediately see that the surface terms break the KW symmetry on the interface and we therefore have to introduce the interface coupling to cancel this term and recover the KW invariance. 
Introducing the same surface term $H_b$ as in 
\eqref{hamiltonianInterface}, we can study the open string sector spectrum as we did for the lattice model in the main text. 
The location of the interface $N_I$ has to be chosen appropriately to compensate for the different effective speeds of light of the Ising lattice model and the TIM Ising model; see \eqref{speedmatch}.
The results are summarized in Fig.~\ref{latticeRG}. 
The crossing point is the conformal interface; one can see that it has exactly the same spectrum of $\Delta$s as the crossing point in Fig.~\ref{tri_Ising} and in fact corresponds to the same Cardy state~\eqref{bondaryansatz}.

\begin{center}
\textbf{Appendix D: Kramers-Wannier invariance of the non-trivial interface }    
\end{center}
In this appendix we verify that our interface is invariant under the KW duality. 
For the Ising model, this was worked out explicitly in Appendix A of~\cite{Aasen:2016dop}.
To do that, we first understand how to study the KW twisted sector in the Ising and tricritical Ising setup. 
The critical step is the Jordan–Wigner transformation
\begin{equation}
\gamma_{2j}=\left(\prod_{k<j} \sigma^x_k\right)\sigma^z_j,\quad 
\gamma_{2j-1}=\left(\prod_{k<j} \sigma^y_k\right)\sigma^z_j \,.
\end{equation}
This leads to 
\begin{equation}
\gamma_{2j}\gamma_{2j+1}=-i Z_jZ_{j+1}\,,\quad
    \gamma_{2j-1}\gamma_{2j}=i X_j\,,
\end{equation}
and 
\begin{equation}
 \gamma_{2j}\gamma_{2j+2}=-iZ_jY_{j+1}\,.
\end{equation}
Therefore, the transverse field Ising Hamiltonian is given by 
\begin{equation}
H_1=-i J\sum_a \gamma_{a}\gamma_{a+1}\,.
\end{equation}
The sub-leading term, which preserves the KW interaction, is now~\cite{OBrien:2017wmx}. 
\begin{equation}
H_2=-K\sum_{a} \gamma_{a-1}\gamma_{a} \gamma_{a+1}\gamma_{a+2}\,.
\end{equation}
The Kramers-Wannier duality, in the fermion basis, corresponds to shifting the location of the fermions  
\begin{equation}
\gamma_{a}\rightarrow\gamma_{a+1}\,.
\end{equation}
To create a KW duality  topological interface, we need to perform the shift for spins on the ``right'' hand side of the interface. That is, to perform the replacement,
\begin{align}
    \gamma_{a}&\rightarrow \gamma_a,\quad {\rm if} \quad a\leq 2k\,,\nonumber\\
    \gamma_{a}&\rightarrow \gamma_{a+1},\quad {\rm if} \quad a> 2k\,.
\end{align}
Notice that this will cause the fermion $\gamma_{2k+1}$ to be decoupled from the rest of the fermions, leading to a two-fold degeneracy on every state of the Hamiltonian. In the original spin basis, we get 
\begin{align}
    H_1&=-J\sum_{i=1}^{k-1}Z_iZ_{i+1}+J\sum_{i=1}^{k}X_{i}
    -JZ_{k}Y_{k+1}-J\sum_{i=k+1}^{N} Z_iZ_{i+1}+J\sum_{i=k+2}^{N}X_{i}\,,
\end{align}
where we have identified site $N+1$ and site $1$. One can easily check that the spectrum of this Hamiltonian  reproduces the KW-twisted sector of the Ising model summarized in Table \ref{IsingKWtwiste}.
To get the tricritical Ising model, we need to add another term,
\begin{align}
    H_2=&-K\sum_{i=1}^{k-2}X_iZ_{i+1}Z_{i+2}-K\sum_{i=1}^{k-2}Z_iZ_{i+1}X_{i+2}\nonumber\\
    &-K X_{k-1}Z_kY_{k+1}+K Z_{k-1}Z_k Z_{k+1}Z_{k+2}+K X_kX_{k+2}+K Z_k Y_{k+1}Z_{k+2}Z_{k+3}\nonumber\\
    &-K\sum_{i=k+2}^{N}X_iZ_{i+1}Z_{i+2}-K\sum_{i=k+1}^{N}Z_iZ_{i+1}X_{i+2}\,.
\end{align}
One can easily check that this reproduces the KW-twisted sector of the tricritical Ising model, summarized in Table~\ref{triIsingKWtwisted}.

\begin{figure}[ht]
\centering
\includegraphics[scale=0.5]{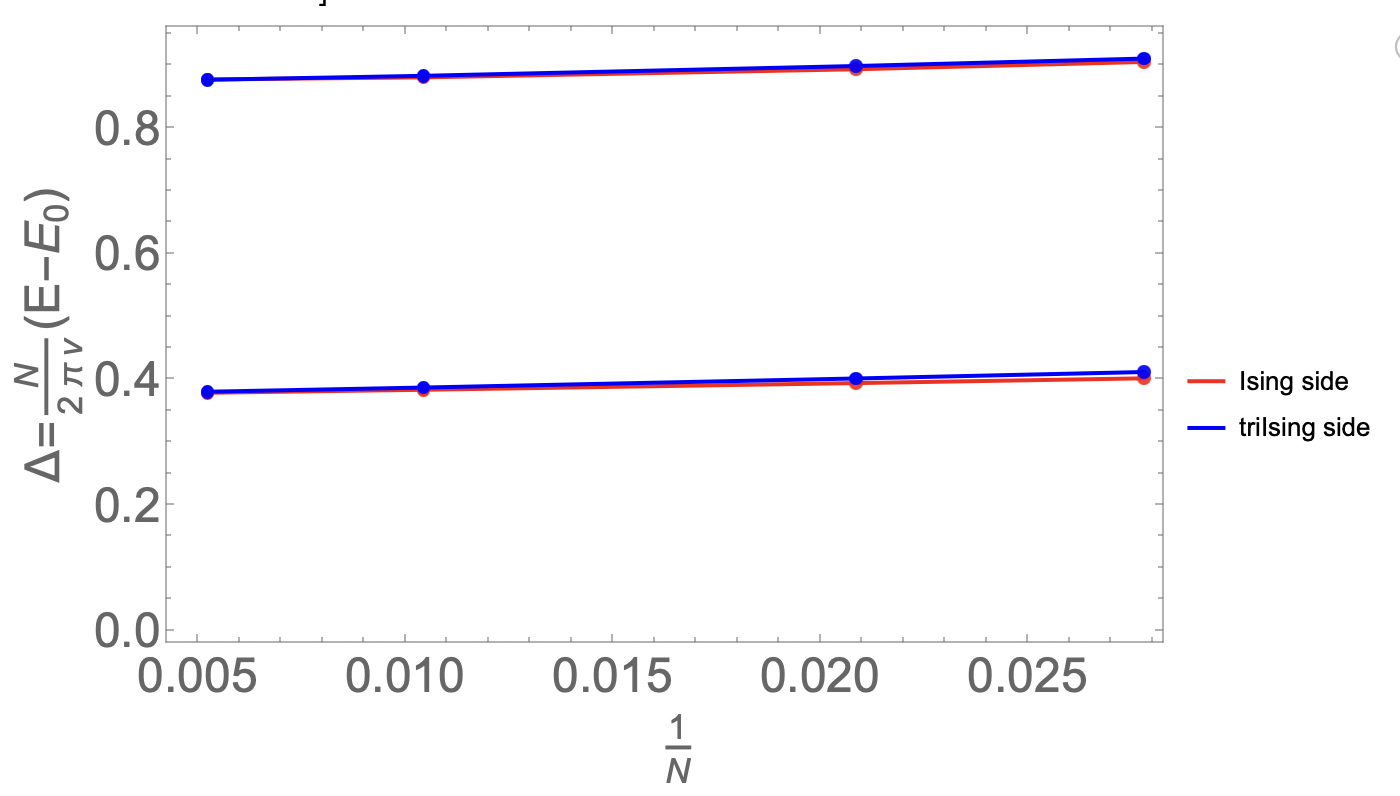}
\caption{The Tri-Ising interface with the KW topological interface on the Ising side and the tricritical Ising side.}\label{KWtwistedinterface}
\end{figure}

\begin{table}[h]
    \centering
    \renewcommand{\arraystretch}{1.5}
    \begin{tabular}{|c|c|c|c|c|}
        \hline
        &  $h_{l}$ & $h_r$ &  $\Delta = h_i + h_j$ &  $s = h_i - h_j$ \\
        \hline
        $(\sigma, \mathbb{I})$ & $1/16$ & $0$ & $1/16$ & $+1/16$ \\
        $(\mathbb{I}, \sigma)$ & $0$ & $1/16$ & $1/16$ & $-1/16$ \\
        $(\sigma, \epsilon)$ & $1/16$ & $1/2$ & $9/16$ & $-7/16$ \\
        $(\epsilon, \sigma)$ & $1/2$ & $1/16$ & $9/16$ & $+7/16$ \\
        \hline
    \end{tabular}
    \caption{Primary spectrum of the Ising CFT in the Kramers-Wannier twisted sector.}\label{IsingKWtwiste}
\end{table}
We can now compare the spectrum of the spin chain by placing the KW topological interface on the Ising and tricritical Ising sides. The result is summarized in Fig.~\ref{KWtwistedinterface}. Since the spectrum matches, we conclude that our interface is invariant under the KW duality transformation.

\begin{table}[h]
    \centering
    \renewcommand{\arraystretch}{1.5}
    \begin{tabular}{|c|c|c|c|c|}
        \hline
          &  $h_l$ &  $h_r$ &  $\Delta = h_i + h_j$ &$s = h_i - h_j$ \\
        \hline
        $(\epsilon, \sigma)$ & $1/10$ & $3/80$ & $11/80$ & $+1/16$ \\
        $(\sigma, \epsilon)$ & $3/80$ & $1/10$ & $11/80$ & $-1/16$ \\
        $(\mathbb{I}, \sigma')$ & $0$ & $7/16$ & $7/16$ & $-7/16$ \\
        $(\sigma', \mathbb{I})$ & $7/16$ & $0$ & $7/16$ & $+7/16$ \\
        $(\epsilon', \sigma)$ & $3/5$ & $3/80$ & $51/80$ & $+9/16$ \\
        $(\sigma, \epsilon')$ & $3/80$ & $3/5$ & $51/80$ & $-9/16$ \\
        $(\epsilon'', \sigma')$ & $3/2$ & $7/16$ & $31/16$ & $+17/16$ \\
        $(\sigma', \epsilon'')$ & $7/16$ & $3/2$ & $31/16$ & $-17/16$ \\
        \hline
    \end{tabular}
    \caption{Primary spectrum of the Tricritical Ising CFT ($c=7/10$) in the Kramers-Wannier twisted sector.}\label{triIsingKWtwisted}
\end{table}

\FloatBarrier

\begin{center}
\textbf{Appendix E: More on the $\mathcal{W}_3$-symmetric Interface }  
\end{center}
In this appendix, we discuss how get $\mathcal{W}_3$ state by gauging the complex conjugation symmetry. This allows us to find the correct open-string channel character. A more rigorous derivation will be given in~\cite{toappear}.

 \medskip 
 
{\bf Gauging complex conjugation} The gauging of the $\mathcal{W}_3$ states described in the text is given by
\begin{align}\label{gauging}
|1\rangle &~\rightarrow | 1, 1 \rangle \nonumber\\
 |D+D^{\dagger}\rangle &~\rightarrow   |L_{-2}; 1,1\rangle \nonumber\\
 i|D-D^{\dagger}\rangle &~\rightarrow   | \epsilon,\epsilon'' \rangle \nonumber\\
 |Z+Z^{\dagger}\rangle &~\rightarrow   | 1,\epsilon' \rangle \nonumber\\
  |Z-Z^{\dagger}\rangle &~\rightarrow  | \epsilon,\epsilon \rangle \nonumber\\
 |X\rangle &~\rightarrow  | L_{-1};\epsilon,\epsilon \rangle \nonumber\\
  |E\rangle &~\rightarrow  | 1,\epsilon \rangle \nonumber\\
|F\rangle &~\rightarrow  | \epsilon, 1 \rangle \nonumber\\
|h=1/10\rangle_{twisted} &~\rightarrow  | \sigma, \sigma \rangle \nonumber\\
|h=1/2\rangle_{twisted} &~\rightarrow  | \sigma, \sigma' \rangle 
\end{align}
where the the states correponding to operators which are odd under charge conjugation are understood to be in the twisted sector of the tensor product theory.
For the states of the $\mathcal{V}irasoro\otimes\mathcal{V}irasoro$ theory, we use the convention $|\phi_{\rm Ising},\phi_{\rm TIM}\rangle$.
The state 
$|L_{-1};\epsilon,\epsilon\rangle$ needs some explanation. 
The state $|\epsilon,\epsilon\rangle$ is a conformal primary of the diagonal $\mathcal{V}irasoro$  algebra. 
At level-1, the state 
$L^{\rm Ising}_{-1}|\epsilon,\epsilon\rangle+ L^{\rm TIM}_{-1}|\epsilon,\epsilon\rangle$
is a descendant. We define 
\begin{equation}
    |L_{-1};\epsilon,\epsilon\rangle \equiv \frac{1}{10}L^{\rm Ising}_{-1}|\epsilon,\epsilon\rangle-\frac{1}{2}L^{\rm TIM}_{-1}|\epsilon,\epsilon\rangle\,.
\end{equation}
It satisfies $(L_{1}^{\rm Ising}+L_{1}^{\rm TIM})|L_{-1};\epsilon,\epsilon\rangle=0$, therefore is a conformal primary. 
The state $|L_{-2};1,1\rangle$ is similarly defined as the level-2 state of the $|1,1\rangle$ $\mathcal{V}irasoro\otimes \mathcal{V}irasoro$ module, which is a conformal primary of the diagonal $\mathcal{V}irasoro$ algebra.
{\bf $\mathcal{W}_3$ Ishibashi states}
There are two types of $\mathcal{W}_3$ Ishibashi states, which can be written as 
\begin{equation}     |h;w\rangle\rangle_+=\sum_{N=0}^{\infty}\sum_{j=1}^{d(N)} |h;w;N,j\rangle_L \otimes {|h;w;N,j\rangle}_R\,,
\end{equation}
or,
\begin{equation}     
|h;w\rangle\rangle_-=\sum_{N=0}^{\infty}\sum_{j=1}^{d(N)} |h;w;N,j\rangle_L \otimes Q {|h;-w;N,j\rangle}_R\,,
\end{equation}
Here $h$ and $w$ are the eigenvalues of $L_0$ and $W_0$ of the primary. $|h,w;N,j\rangle$ is the $j$-th state at level $N$.
Notice that both $L_{-k}$ and $W_{-k}$ raise the level by $k$. 
Also, $d(N)$ is the number of states at level-$N$. 
The state $Q{|h;-w;N,j\rangle}_R$ is identical to the state ${|h;-w;N,j\rangle}_R$ up to flipping the sign of $\mathcal{W}_3$ generators, $W_{-k}\rightarrow-W_{-k}$.
They satisfy
\begin{align}
(L^{(L)}_{n}-{L}^{(R)}_{-n})|h,w\rangle\rangle_{\pm}=0\,,\quad (W^{(L)}_{n}\mp{W}^{(R)}_{-n})|h,w\rangle\rangle_{\pm}=0\,.
\end{align}
The twisted Ishibashi state $|h,w\rangle\rangle_{-}$ plays a crucial role in finding the extra boundary conditions of the 3-state Potts model beyond Cardy's solutions~\cite{Affleck:1998nq}. 
Acting on the level-zero state, we have 
\begin{align}
    (W^{(L)}_{n}\mp{W}^{(R)}_{-n})|h,w\rangle_{L}\otimes|\bar{h},\bar{w}\rangle_{R}=0\,.
\end{align}
The condition requires that $w+\bar{w}=0$. 
If we were to study the boundary conditions of $\mathcal{W}_3$, this is not allowed unless $w=\bar{w}=0$, since the bulk states require $w=\bar{w}$.
We are, however, studying the boundary condition of the $\mathcal{V}irasoro\otimes\mathcal{V}irasoro$ theory, twisted Ishibashi states with non-zero $w$ are also allowed. 
In particular, the leading field of the Ishibashi state $\left(\frac{1}{\sqrt{2}}|D\rangle\rangle_{\mathcal{W}_3,-}+\frac{1}{\sqrt{2}} |D^{\dagger}\rangle\rangle_{\mathcal{W}_3,-}\right)$ is 
\begin{eqnarray}
&&\frac{1}{\sqrt{2}} |D\rangle_L\otimes|D^{\dagger}\rangle_R+\frac{1}{\sqrt{2}} |D^{\dagger}\rangle_L\otimes|D\rangle_R\nonumber\\&=&\frac{1}{\sqrt{2}}\left(|D+D^{\dagger}\rangle_L\otimes |D+D^{\dagger}\rangle_R-|D-D^{\dagger}\rangle_L\otimes |D-D^{\dagger}\rangle_R\right)\nonumber\\
&\xrightarrow{\rm gauging}& |L_{-2}; 1,1\rangle_L\otimes|L_{-2}; 1,1\rangle_R- | \epsilon,\epsilon'' \rangle_L\otimes |\epsilon,\epsilon'' \rangle_R.
\end{eqnarray}
The untwisted Ishibashi state $\left(\frac{1}{\sqrt{2}}|D\rangle\rangle_{\mathcal{W}_3,+}+\frac{1}{\sqrt{2}} |D^{\dagger}\rangle\rangle_{\mathcal{W}_3,+}\right)$, on the other hand, has the leading field
\begin{eqnarray}
&&\frac{1}{\sqrt{2}} |D\rangle_L\otimes|D\rangle_R+\frac{1}{\sqrt{2}} |D^{\dagger}\rangle_L\otimes|D^{\dagger}\rangle_R\nonumber\\&=&\frac{1}{\sqrt{2}}\left(|D+D^{\dagger}\rangle_L\otimes |D+D^{\dagger}\rangle_R+|D-D^{\dagger}\rangle_L\otimes |D-D^{\dagger}\rangle_R\right)\nonumber\\
&\xrightarrow{\rm gauging}& |L_{-2}; 1,1\rangle_L\otimes|L_{-2}; 1,1\rangle_R+ | \epsilon,\epsilon'' \rangle_L\otimes |\epsilon,\epsilon'' \rangle_R.
\end{eqnarray}

{\bf Relation between $\mathcal{W}_3$ and Vir$\times$Vir}

There are relations between the $\mathcal{W}_3$ characters and the $\mathcal{V}irasoro\otimes \mathcal{V}irasoro$ characters: 
\begin{align}
\chi_{h=0}^{\mathcal{W}_3}(q) + 2\chi_{h=2}^{\mathcal{W}_3}(q) &= \chi_1^{\text{Ising}}(q)\chi_1^{\text{TIM}}(q) + \chi_{\epsilon}^{\text{Ising}}(q)\chi_{\epsilon''}^{\text{TIM}}(q) \nonumber\\
2\chi_{h=3/5}^{\mathcal{W}_3}(q) + \chi_{h=8/5}^{\mathcal{W}_3}(q) &= \chi_1^{\text{Ising}}(q)\chi_{\epsilon'}^{\text{TIM}}(q) + \chi_{\epsilon}^{\text{Ising}}(q)\chi_{\epsilon}^{\text{TIM}}(q) \nonumber\\
\chi_{h=1/10}^{\mathcal{W}_3}(q) &= \chi_{\sigma}^{\text{Ising}}(q)\chi_{\sigma}^{\text{TIM}}(q) \nonumber\\
\chi_{1/2}^{\mathcal{W}_3}(q) &= \chi_{\sigma}^{\text{Ising}}(q)\chi_{\sigma'}^{\text{TIM}}(q). \nonumber\\
\chi_{h=1/10}^{\mathcal{W}_3}(q) &= \chi_1^{\text{Ising}}(q)\chi_{\epsilon}^{\text{TIM}}(q) + \chi_{\epsilon}^{\text{Ising}}(q)\chi_{\epsilon'}^{\text{TIM}}(q) \nonumber\\
\chi_{h=1/2}^{\mathcal{W}_3}(q) &= \chi_1^{\text{Ising}}(q)\chi_{\epsilon''}^{\text{TIM}}(q) + \chi_{\epsilon}^{\text{Ising}}(q)\chi_1^{\text{TIM}}(q)\,.
\end{align}
There are similar relations among the Ishibashi states
\begin{align}
|1\rangle\rangle_{\mathcal{W}_3,+}+|D\rangle\rangle_{\mathcal{W}_3,+}+|D^{\dagger}\rangle\rangle_{\mathcal{W}_3,+}&=|1,1\rangle\rangle_{\mathcal{V}ir\otimes \mathcal{V}ir}+|\epsilon,\epsilon''\rangle\rangle_{\mathcal{V}ir\otimes \mathcal{V}ir} \nonumber\\
|Z\rangle\rangle_{\mathcal{W}_3,+}+|Z^{\dagger}\rangle\rangle_{\mathcal{W}_3,+}+|X\rangle\rangle_{\mathcal{W}_3,+}&=|1,\epsilon'\rangle\rangle_{\mathcal{V}ir\otimes \mathcal{V}ir}+|\epsilon,\epsilon\rangle\rangle_{\mathcal{V}ir\otimes \mathcal{V}ir}\nonumber\\
|F\rangle\rangle_{\mathcal{W}_3,+}&=|\epsilon,1\rangle\rangle_{\mathcal{V}ir\otimes \mathcal{V}ir}+|1,\epsilon''\rangle\rangle_{\mathcal{V}ir\otimes \mathcal{V}ir}\nonumber\\
|E\rangle\rangle_{\mathcal{W}_3,+}&=|1,\epsilon \rangle\rangle_{\mathcal{V}ir\otimes \mathcal{V}ir}+|\epsilon,\epsilon'\rangle\rangle_{\mathcal{V}ir\otimes \mathcal{V}ir}\,.\nonumber
\end{align}
The Ising$\otimes$TIM CFT has 18 Cardy states, among them, two boundary states preserve the $\mathcal{W}_3$ symmetry:
\begin{align}
|C_1\rangle\equiv&|h=1/16\rangle_{\rm Cardy}^{\rm Ising} \otimes |h=7/16\rangle_{\rm Cardy}^{\rm TIM}\nonumber\\=&\sqrt{c_3}\left(|1\rangle\rangle_{\mathcal{W}_3,+}+|D\rangle\rangle_{\mathcal{W}_3,+}+|D^{\dagger}\rangle\rangle_{\mathcal{W}_3,+}\right)+\sqrt{c_2}\left(|Z\rangle\rangle_{\mathcal{W}_3,+}+|Z^{\dagger}\rangle\rangle_{\mathcal{W}_3,+}+|X\rangle\rangle_{\mathcal{W}_3,+}\right)\nonumber\\&-\sqrt{c_3}|F\rangle\rangle_{\mathcal{W}_3,+}-\sqrt{c_2}|E\rangle\rangle_{\mathcal{W}_3,+}\nonumber\\
|C_2\rangle\equiv&|h=1/16\rangle_{\rm Cardy}^{\rm Ising} \otimes |h=3/80\rangle_{\rm Cardy}^{\rm TIM}\nonumber\\
=&\sqrt{c_4}\left(|1\rangle\rangle_{\mathcal{W}_3,+}+|D\rangle\rangle_{\mathcal{W}_3,+}+|D^{\dagger}\rangle\rangle_{\mathcal{W}_3,+}\right)-\sqrt{c_5}\left(|Z\rangle\rangle_{\mathcal{W}_3,+}+|Z^{\dagger}\rangle\rangle_{\mathcal{W}_3,+}+|X\rangle\rangle_{\mathcal{W}_3,+}\right)\nonumber\\&-\sqrt{c_4}|F\rangle\rangle_{\mathcal{W}_3,+}+\sqrt{c_5}|E\rangle\rangle_{\mathcal{W}_3,+}\,.\nonumber
\end{align}
We have defined
$$c_3=\sqrt{\frac{1}{10} \left(5-\sqrt{5}\right)},~ c_4=\sqrt{1+\frac{2}{\sqrt{5}}},~c_5=\sqrt{1-\frac{2}{\sqrt{5}}}\,.$$

{\bf More on the mixed-boundary partition function}
We can study the annulus partition with the ${}_{\rm FM}\langle B|$ boundary condition imposed on one side of the annulus, and $|C_i\rangle_{}$ boundary condition imposed on the other side of the annulus.
Since off-diagonal amplitudes in the closed-string channel must admit a positive-integer character decomposition in the open-string channel, we find that the only consistent choice is to fix $s_3=-1$, and choose the twisted Ishibashi states in \eqref{bondaryansatz}.
The partition function is then becomes
\begin{equation} 
    Z(q)_{B, C_1}= {}_{\rm FM}\langle B| e^{-\beta H} |C_1\rangle_{}=\sqrt{c_1c_3}\tilde{\chi}^{\mathcal{W}_3}_{h=0}(q)-\sqrt{2}c_2\tilde{\chi}^{\mathcal{W}_3}_{h=8/5}(q),
\end{equation}
and 
\begin{equation}
    Z(q)_{B, C_2}= {}_{\rm FM}\langle B| e^{-\beta H} |C_2\rangle_{}=\sqrt{c_1c_4}\tilde{\chi}^{\mathcal{W}_3}_{h=0}(q)+\sqrt{2c_2c_5}\tilde{\chi}^{\mathcal{W}_3}_{h=8/5}(q).
\end{equation}
Here $\tilde{\chi}_{h=0}(q)$ and $\tilde{\chi}_{h=8/5}(q)$ are the twisted $\mathcal{W}_3$ characters. They are given by normal $\mathcal{W}_3$ characters, but flipping the signs of the terms corresponding to the states created by an odd number of $\mathcal{W}_3$ generates. 
After the S-transformation, they become 
\begin{align}
    Z(q)_{B, C_1}&=\tilde{\chi}^{A_2^{(2)}}_{h=1/16}(\tilde{q})+\tilde{\chi}^{A_2^{(2)}}_{h=7/16}(\tilde{q}),  \nonumber\\
       Z(q)_{B, C_2}&=\tilde{\chi}^{A_2^{(2)}}_{h=3/80}(\tilde{q})+\tilde{\chi}^{A_2^{(2)}}_{h=13/80}(\tilde{q}).
\end{align}
We have used the S-transformation matrix of the twisted $\mathcal{W}_3$ characters calculated based on~\cite{Kac_1984}, which are
\begin{equation} 
\left(
\begin{array}{c}
 \tilde{\chi }^{A_2^{(2)}}_{h=3/80}(\tilde{q}) \\
 \tilde{\chi }^{A_2^{(2)}}_{h=13/80}(\tilde{q})  \\
 \tilde{\chi }^{A_2^{(2)}}_{h=7/16}(\tilde{q})  \\
 \tilde{\chi }^{A_2^{(2)}}_{h=1/16}(\tilde{q})  \\
\end{array}
\right)=\left(
\begin{array}{cccc}
 \frac{1}{2} \sqrt{1+\frac{1}{\sqrt{5}}} & \frac{1}{2} \sqrt{1+\frac{1}{\sqrt{5}}} & \frac{1}{2} \sqrt{1-\frac{1}{\sqrt{5}}} & \frac{1}{2} \sqrt{1-\frac{1}{\sqrt{5}}} \\
 \frac{1}{2} \sqrt{1+\frac{1}{\sqrt{5}}} & -\frac{1}{2} \sqrt{1+\frac{1}{\sqrt{5}}} & \frac{1}{2} \sqrt{1-\frac{1}{\sqrt{5}}} & -\frac{1}{2} \sqrt{1-\frac{1}{\sqrt{5}}} \\
 \frac{1}{2} \sqrt{1-\frac{1}{\sqrt{5}}} & \frac{1}{2} \sqrt{1-\frac{1}{\sqrt{5}}} & -\frac{1}{2} \sqrt{1+\frac{1}{\sqrt{5}}} & -\frac{1}{2} \sqrt{1+\frac{1}{\sqrt{5}}} \\
 \frac{1}{2} \sqrt{1-\frac{1}{\sqrt{5}}} & -\frac{1}{2} \sqrt{1-\frac{1}{\sqrt{5}}} & -\frac{1}{2} \sqrt{1+\frac{1}{\sqrt{5}}} & \frac{1}{2} \sqrt{1+\frac{1}{\sqrt{5}}} \\
\end{array}
\right).\left(
\begin{array}{c}
 \tilde{\chi }_{h=0}(q) \\
 \tilde{\chi }_{h=1/2}(q)  \\
 \tilde{\chi }_{h=8/5}(q)  \\
 \tilde{\chi }_{h=1/10}(q)  \\
\end{array}
\right).
\end{equation}
The affined twisted $A_2^{(2)}$ characters can be calculated following Rocha-Caridi's approach to subtract the character of submodules generated by these null states, see~\cite{Kac_1984,RochaCaridi}. 
As a power series, they are
\begin{align}
 \tilde{\chi }^{A_2^{(2)}}_{h=3/80}(q)&=q^{3/80}+q^{43/80}+q^{83/80}+2 q^{123/80}+3 q^{163/80}+4 q^{203/80}+5 q^{243/80}+7 q^{283/80}+\ldots,\nonumber\\
  \tilde{\chi }^{A_2^{(2)}}_{h=13/80}(q)&=q^{13/80}+q^{53/80}+2 q^{93/80}+2 q^{133/80}+3 q^{173/80}+4 q^{213/80}+6 q^{253/80}+7 q^{293/80}+\ldots,\nonumber\\
    \tilde{\chi }^{A_2^{(2)}}_{h=7/16}(q) &=q^{7/16}+q^{15/16}+q^{23/16}+2 q^{31/16}+2 q^{39/16}+3 q^{47/16}+4 q^{55/16}+5 q^{63/16}+\ldots,\nonumber\\
    \tilde{\chi }^{A_2^{(2)}}_{h=1/16}(q) &= q^{1/16}+q^{17/16}+q^{25/16}+2 q^{33/16}+2 q^{41/16}+4 q^{49/16}+4 q^{57/16}+\ldots. 
\end{align}
Notice that these open-string channel states have $\Delta$ values that are half-integer-spaced. 
This has a clear explanation.
\begin{figure}[t]
\centering
\includegraphics[scale=0.7]{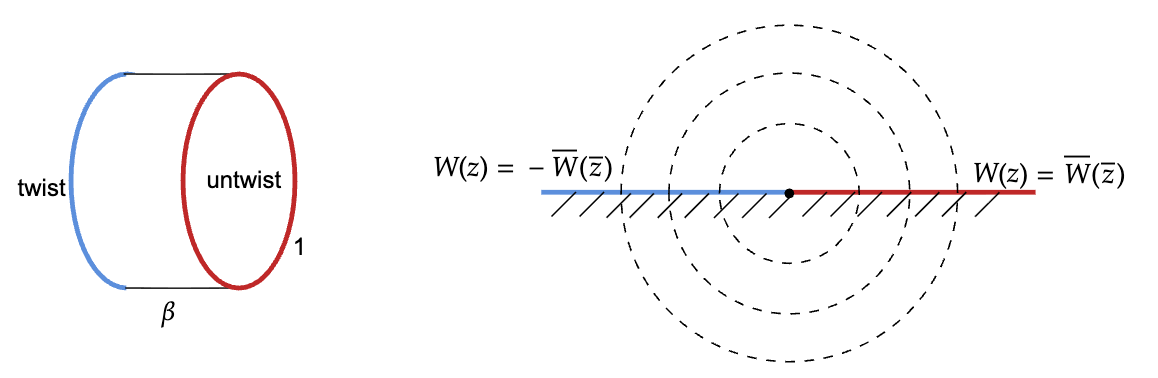}
\caption{The mixed boundary partition function. The spin-3 current satisfies different boundary conditions on the two sides.}\label{spin3twist}
\end{figure}
The spin-3 current satisfies the untwisted boundary condition on one side, and the twisted boundary condition on the other side.
Only half-integer modes
$$
W_{-1/2-k}
$$
are allowed, see Fig.~\ref{spin3twist}. 
Together with the integer modes of the Virasoro generators, they lead to the half-integer-spaced spectra. 

{\bf More $\mathcal{W}_3$ invariant states}

Since the spin-3 current \eqref{spin3current} is invariant under the Fibonacci symmetry, we can build two more states $\mathcal{T}_{\tau}|B\rangle_{\rm FM/AFM}$ which preserve the $\mathcal{W}_3$ symmetry. 
One can also check that 
\begin{align}
{}_{\rm FM}\langle B| e^{-\beta H} \mathcal{T}_{\tau}|B\rangle_{\rm FM/AFM}=&~\chi^{\mathcal{W}_3}_{h=1/10}(\tilde{q})+2\chi^{\mathcal{W}_3}_{h=3/5}(\tilde{q})+\chi^{\mathcal{W}_3}_{h=8/5}(\tilde{q})\nonumber\\
&+\chi^{\mathcal{W}_3}_{h=2/45}(\tilde{q})+\chi^{\mathcal{W}_3}_{h=17/45}(\tilde{q})+\chi^{\mathcal{W}_3}_{h=32/45}(\tilde{q})\,.
\end{align}
can be decomposed into $\mathcal{W}_3$ character with positive integer coefficients in the open-string channel. 
Here $\mathcal{T}_{\tau}$ stands for a Fibonacci topological line along the temporal direction.

\end{document}